\documentclass[pra,aps,twocolumn,showpacs]{revtex4} 
\usepackage{graphicx} 
\newcommand{\ket}[1]{|#1\rangle} 
\newcommand{\bra}[1]{\langle #1|} 
 
\begin{document}
\title{Geometric phase in weak measurements}
\author{Erik Sj\"{o}qvist\footnote{Electronic address: 
eriks@kvac.uu.se}} 
\affiliation{Department of Quantum Chemistry, Uppsala University, 
Box 518, Se-751 20 Uppsala, Sweden}
\date{\today}       
\begin{abstract}
Pancharatnam's geometric phase is associated with the phase 
of a complex-valued weak value arising in a certain type of 
weak measurement in pre- and post-selected quantum ensembles. 
This makes it possible to test the nontransitive nature of 
the relative phase in quantum mechanics, in the weak measurement 
scenario. 
\end{abstract}
\maketitle
Pancharatnam's \cite{pancharatnam56} notion of `in-phase' is 
the underlying principle of the quantal geometric phase 
\cite{ramaseshan86,berry87}. It is based on the physical 
condition of maximal interference: the vector representatives
$\ket{A}$ and $\ket{B}$ of two nonorthogonal quantal states $A$ 
and $B$ are in-phase if they produce a maximum in intensity 
when added. This amounts to the condition $\bra{B}A\rangle >0$, 
and any deviation from it defines the Pancharatnam relative 
phase $\phi_{BA}$ as
\begin{eqnarray}
e^{i\phi_{BA}} = \frac{\bra{B}A\rangle}{|\bra{B}A\rangle|} ,   
\end{eqnarray}
i.e., the phase of the complex-valued scalar product $\bra{B}A\rangle$. 

The Pancharatnam relative phase is a nontransitive concept in the
sense that it is possible to have $\phi_{BA} \neq 0$ although 
both $\phi_{CB}$ and $\phi_{AC}$ vanish. Moreover, if $\phi_{CB} = 
\phi_{AC}=0$ then the excess phase $\phi_{BA}$ equals the geometric 
phase 
\begin{eqnarray}
\Delta (A,B,C) = 
\arg \langle A \ket{C} \langle C \ket{B} \langle B \ket{A} 
\end{eqnarray} 
for the geodesic triangle connecting the quantal states $A,B$, and $C$
in projective Hilbert space. For a qubit, such as the polarisation of
a photon, $\Delta (A,B,C)$ is proportional to the solid angle of the
spherical triangle on the Bloch sphere with vertices at $A,B$, and
$C$. 

The triangle phase $\Delta$ is the smallest nontrivial entity
of the geometric phase. To see this, let $A_1,A_2,\ldots,A_n$ be an
arbitrary sequence $c$ of pure quantal states. The geometric phase
$\gamma [c]$ associated with this sequence takes the form
\begin{eqnarray} 
\gamma [c] = \arg \bra{A_1} A_n \rangle \bra{A_n} A_{n-1} \rangle 
\ldots \bra{A_2} A_1 \rangle .  
\end{eqnarray}
Using elementary properties of complex numbers, one may rewrite 
$\gamma [c]$ as \cite{mukunda93}
\begin{eqnarray} 
\gamma [c] = \sum_{k=2}^{n-1} \Delta (A_1,A_k,A_{k+1}) . 
\end{eqnarray}
In the continuous limit, this takes the form 
\begin{eqnarray} 
\gamma [c] & = & \int_0^T \Delta (A_0,A_t,A_{t+dt}) 
\nonumber \\ 
 & = & \arg \bra{A_0} A_T \rangle + 
i \int_0^T \bra{A_t} \dot{A}_t \rangle dt ,  
\end{eqnarray}
where $t\in [0,T] \rightarrow A_s$ is a smooth path in projective 
Hilbert space. If $\ket{A_0}$ and $\ket{A_T}$ are in-phase, then 
$\gamma [c]$ reduces to the definition of geometric phase in 
\cite{aharonov87,samuel88}. 

Here, we propose a quantitative test of the nontransitive nature
of Pancharatnam's concept of relative phase by utilising weak 
measurements \cite{aharonov90}. It measures $\Delta (A,B,C)$ in 
terms of the phase of a complex-valued weak value \cite{aharonov90}
obtained in a weak measurement of a one-dimensional projector 
performed between two complete projective measurements.

The standard techniques to measure the geometric phase are based 
on interferometric \cite{wagh95a} or polarimetric ideas
\cite{wagh95b}. Common to these ideas is their need for phase
calibration, which constitutes an extra constraint in the experimental
set up. Curiously, as shall be clear from the following analysis,
phase calibration seems not needed in the weak measurement scenario.

Let us first delineate in what sense polarimetry and interferometry 
involve phase calibration. In polarimetry, an additional variable 
U(1) shift $e^{i\chi}$ is applied to $\ket{A}$ or $\ket{B}$, when 
superposed. The resulting state is thereafter projected onto the 
third state $C$, yielding the intensity 
\begin{eqnarray} 
\mathcal{I}_p \propto 1 + \textrm{Re} \left( e^{-i\chi} 
\langle A \ket{C} \langle C \ket{B} \right) ,  
\label{eq:pol}
\end{eqnarray} 
where we have assumed that $A$ and $B$ have equal probability weight
and that $e^{i\chi}$ is applied to $\ket{A}$. The shift in the 
interference oscillations equals $\Delta (A,B,C)$ provided $\chi$ is
set to zero at $\phi_{AB}$. Thus, to measure the geometric phase we
must first obtain the relative phase between $\ket{A}$ and $\ket{B}$,
e.g., by measuring the intensity $\big|
\ket{A} + \ket{B}\big|^2 \propto 1+|\langle A \ket{B}|
\cos \phi_{AB}$, and thereafter shift the interference pattern 
in Eq. (\ref{eq:pol}) by $\phi_{AB}$. 

A more symmetric expression than Eq. (\ref{eq:pol}) 
is obtained in interferometry, where an input internal state $A$ is
distributed on two spatial particle beams. One of the beams is
exposed to a sequence of projections $\ket{A} \rightarrow
\ket{B} \langle B\ket{A} \rightarrow \ket{C}
\langle C \ket{B} \langle B\ket{A}$, while the other beam is exposed 
to the variable U(1) phase shift $e^{i\chi}$. The resulting intensity 
oscillations 
\begin{eqnarray} 
\mathcal{I}_i \propto 1 + \textrm{Re} \left( e^{-i\chi} 
\langle A \ket{C} \langle C \ket{B} \langle B \ket{A} \right) 
\end{eqnarray} 
are shifted by the geometric phase $\Delta (A,B,C)$ provided 
$\chi$ is kept to zero relative a reference interferometer 
defined by removing the projections. Physically, this means that 
the relative path length of the two interferometer arms are kept 
equal throughout the experimental run.   

Now, we describe the weak measurement approach to measure the
geometric phase $\Delta (A,B,C)$. Let a quantum system and measurement
device be prepared in the product state
\begin{eqnarray} 
\varrho_0 = \ket{A} \bra{A} \otimes \ket{M_0} \bra{M_0} .   
\end{eqnarray}
We take the initial apparatus wave function in the position representation 
as a Gaussian. Explicitly, we may take $M_0(q) = \langle q 
\ket{M_0} \sim e^{-q^2/(2\sigma^2)}$, where $\ket{q}$ is an eigenvector 
of the pointer position operator $Q$. The system and device are made to 
interact during a finite time interval as described by the Hamiltonian 
\cite{vonneumann55}
\begin{eqnarray} 
H(t) = g(t) \mathcal{P}^B \otimes Q .   
\end{eqnarray}
Here, $\mathcal{P}^B$ is assumed to be a one-dimensional projector, i.e., 
$\mathcal{P}^B = \ket{B} \bra{B}$, and the time-dependent coupling 
parameter $g$ turns on and off the interaction between the 
measurement device and the measured system.  
When the interaction is over, we obtain the total state 
\begin{eqnarray} 
\varrho & = & e^{-i\kappa  \mathcal{P}^B \, \otimes \, Q} 
\ket{A} \bra{A} \otimes \ket{M_0} \bra{M_0} 
e^{i\kappa \mathcal{P}^B \, \otimes \, Q}        
\end{eqnarray}
with the measurement strength 
\begin{eqnarray}
\kappa = \frac{1}{\hbar} \int g (t) dt ,   
\end{eqnarray} 
where $2\pi \hbar$ is Planck's constant. Conditioned on the 
post-selection of the state $\ket{C} \bra{C}$, we find
\begin{eqnarray} 
\varrho_{ps} & = & \ket{C} \bra{C} \otimes 
\bra{C} e^{-i\kappa  \mathcal{P}^B \, \otimes \, Q} \ket{A} \ket{M_0} 
\nonumber \\ 
 & & \times \bra{M_0} \bra{A} e^{i\kappa \mathcal{P} ^B \, \otimes \, Q} 
\ket{C} .        
\end{eqnarray}
Now, if $\kappa \sigma \ll 1$ then terms in the order of $(\kappa
Q)^n$, $n\geq 2$, can be neglected. Under this condition and the 
assumption that $\langle A\ket{C} \neq 0$, we obtain 
\begin{eqnarray}
\bra{C} e^{-i\kappa  \mathcal{P}^B \, \otimes \, Q} \ket{A} \approx 
\bra{C}A\rangle e^{-i\kappa  \mathcal{P}_w^B (A,C) \, Q} , 
\end{eqnarray}
where 
\begin{eqnarray} 
\mathcal{P}_w^B (A,C) & = & 
\frac{\bra{C} \mathcal{P}^B \ket{A}}{\langle C \ket{A}} = 
\frac{\langle C \ket{B} \langle B \ket{A}}{\langle C \ket{A}}
\label{eq:weakvalue} 
\end{eqnarray}
is the weak value of the operator $\mathcal{P}^B$ with respect to the
pre-selected state $A$ and the post-selected state $C$. The resulting
wave packet of the measuring device reads
\begin{eqnarray} 
M (q) & \sim & 
\exp \left( - \frac{(q-\kappa \sigma^2 \textrm{Im} 
\mathcal{P}_w^B (A,C))^2}{2\sigma^2} \right) 
\nonumber \\ 
 & & \times \exp \Big( -i\kappa \textrm{Re} 
\mathcal{P}_w^B (A,C) q \Big) .  
\end{eqnarray}
Thus, as a result of the post-selection, the weak measurement shifts 
the position of the pointer by 
\begin{eqnarray} 
\delta q = \kappa \sigma^2 \textrm{Im} 
\mathcal{P}_w^B (A,C)
\label{eq:q}
\end{eqnarray} 
and its momentum by 
\begin{eqnarray} 
\delta p = -\hbar \kappa 
\textrm{Re} \mathcal{P}_w^B (A,C). 
\label{eq:p}
\end{eqnarray} 
Upon multiplication and division of the right-hand side of Eq. 
(\ref{eq:weakvalue}) by $\langle A \ket{C}$, we obtain the relation 
\begin{eqnarray} 
\arg \mathcal{P}_w^B (A,C) = \Delta (A,B,C) , 
\end{eqnarray}
which, together with Eqs. (\ref{eq:q}) and (\ref{eq:p}), yields the 
main result 
\begin{eqnarray} 
\Delta (A,B,C) = 
-\arctan \left( \frac{\hbar}{\sigma^2} \frac{\delta q}{\delta p} \right) . 
\end{eqnarray}
Thus, by measuring the shift in position and momentum of the 
measuring device immediately after turning off the weak interaction, 
we obtain the Pancharatnam geometric phase. Clearly, this deduction   
is independent of any phase relation between the pre- and post-selected 
vectors $\ket{A}$ and $\ket{C}$. 

The above scheme can be used to measure the geometric phase for any
polygon-shaped path $A_1 ,A_2 \ldots A_n$ in projective Hilbert
space. Prepare $A_1$, measure $\mathcal{P}^{A_k} = \ket{A_k}
\bra{A_k}$ weakly, and post-select $A_{k+1}$. Adding the phases of 
the resulting weak values $\mathcal{P}_w^{A_k} (A_1,A_{k+1})$, 
$k=2,\ldots,n-1$, yields $\Delta (A_1,A_2,\ldots,A_n)$.

\begin{figure}[htb]
\includegraphics[width = 8.5cm]{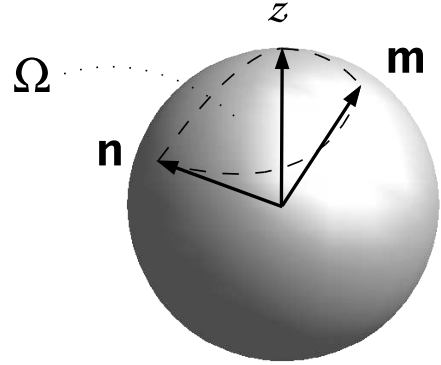}
\caption{\label{fig:sphere}Bloch sphere representation of weak spin 
measurement that yields the geometric phase. ${\bf n}$ and ${\bf m}$ 
are the pre- and post-selected states. The solid angle $\Omega$ 
can be inferred by intermediate weak measurement of the spin in 
the $z$ direction.} 
\end{figure}

A physically reasonable scenario for the weak measurement procedure 
could be the following variant of the Stern-Gerlach experiment in 
\cite{aharonov88}. Let the incoming beam of spin$-\frac{1}{2}$ 
particles be fully polarised in a direction represented by the 
unit vector ${\bf n}$ in three-dimensional space. The spin is 
projected weakly onto the $\uparrow_z$ state, followed by
post-selection in a direction represented by another unit 
vector ${\bf m}$. The resulting weak value becomes
\begin{eqnarray}
\uparrow_{z;w} & = & 
\frac{\sqrt{\big( 1+n_z+m_z+{\bf n}\cdot{\bf m} \big)^2 + 
\big( [{\bf n}\times{\bf m}]_z \big)^2}}
{2\big( 1+{\bf n}\cdot{\bf m}\big)} 
\nonumber \\ 
 & & \times \exp \left( -i\frac{1}{2}\Omega \right) , 
\label{eq:spin}
\end{eqnarray} 
where $\Omega$ is the solid angle of the geodesic triangle on 
the Bloch sphere with vertices at ${\bf n}$, $z$, and ${\bf m}$, 
see Fig. \ref{fig:sphere}. It follows that $\arg \uparrow_{z;w} = 
-\frac{1}{2} \Omega$, which is the expected geometric phase 
for the geodesic triangle \cite{pancharatnam56,ramaseshan86,berry87}. 

As is apparent from Eq. (\ref{eq:spin}), the weak value
$\uparrow_{z;w}$ diverges when ${\bf m} \rightarrow -{\bf n}$. The
reason for this behavior is that for a system pre-selected at the
point ${\bf n}$, there is, in the weak measurement limit, a vanishing
probability for it to pass a post-selection of $-{\bf n}$ \cite{remark}. 
Now, the antipodal points $\pm{\bf n}$ precisely correspond to
orthogonal states for which the geometric phase is undefined. In 
other words, the case where the geometric phase becomes undefined
corresponds to a post-selected quantum ensemble whose size tend to 
zero in the weak measurement limit.

In conclusion, we have put forward a procedure for measuring
Pancharatnam's geometric phase in certain weak measurements in pre-
and post-selected quantum ensembles. The proposed test avoids any 
need for a phase relation between the pre- and post-selected states. 
Implementation of the weak measurement scheme for observing 
the Pancharatnam geometric phase should be feasible for 
neutron spin \cite{golub89a,golub89b} or photon polarisation 
\cite{knight90,pryde05}.
\vskip 0.3 cm 
Financial support from the Swedish Research Council is acknowledged. 
   

\begin{thebibliography}{99}
\bibitem{remark} Close to the antipodal point, the magnitude 
of the weak value can be very large \cite{aharonov88}.
\end{thebibliography}
\end{document}